\documentclass[12pt]{article}
 
\usepackage{epsfig,float}

\newcommand{\KS}{\mbox{$K_S^0$}}

\newcommand{\acp}	{\mbox{${{\cal A}_{\rm CP}}$}}

\newcommand{\sig}	{\mbox{${ \cal S   }$}}
\newcommand{\sigbar}	{\mbox{${ \overline{\sig} }$}}
\newcommand{\bkg}	{\mbox{${ \cal B   }$}}
\newcommand{\bkgbar}	{\mbox{${ \overline{\bkg} }$}}

\newcommand{\pdf}  	{\mbox{${PDF}$}}
\newcommand{\kp}	{\mbox{$K^\pm\pi^\mp$}}
\newcommand{\op}	{\mbox{$\omega\pi^\pm$}}

\newcommand{\kpz}	{\mbox{$K^+\pi^0$}}
\newcommand{\ppz}	{\mbox{$\pi^+\pi^0$}}

\newcommand{\ek}	{\mbox{$\eta^\prime K^\pm$}}
\newcommand{\epp}	{\mbox{$\eta\pi^+\pi^-$}}
\newcommand{\rg}	{\mbox{$\rho\gamma$}}

\newcommand{\dedx}	{\mbox{$dE/dx$}}

\newcommand{\gev}	{\mbox{${\rm ~GeV}$}}

\newcommand{\etapr}	{\mbox{${\eta^\prime}$}}

\newcommand{\qqb}	{\mbox{$q\bar q$}}
\newcommand{\bbar}	{\mbox{$\bar b$}}

\newcommand{\fbar}	{\mbox{$\bar f$}}

\newcommand{\eeqq}	{\mbox{$ee\to\qqb$}}

\newcommand{\branch}    {\mbox{${Br}$}}

\newcommand{\lum}    	{\mbox{${\cal L}$}}
\newcommand{\like}    	{\mbox{${\cal L}$}}

\newcommand{\cossph}	{\mbox{$\cos \theta_{\rm sph}$}}

\newcommand{\de}	{\mbox{$\Delta E$}}
\newcommand{\mb}	{\mbox{$M_B$}}

\newcommand{\aerr}[3]   {\mbox{${{#1}^{+ #2}_{- #3}}$}}

\begin{document}
\vskip 1.0cm
\rightline{CLEO CONF 99-16}
\vspace{2.0cm}
\begin{center}
\begin{Large}
{\bf Measurement of Charge Asymmetries in Charmless Hadronic $B$ Decay}
\end{Large}
\vskip 1.0cm
{\large CLEO Collaboration}
\vskip 0.4 cm
(August 6th, 1999)
\end{center}
\vspace{0.5cm}

\begin{abstract}
      We search for CP violating asymmetries in the charmless hadronic
      $B$ meson decays to
      $ K^\pm \pi^\mp$,
      $ K^\pm \pi^0$,
      $ \KS \pi^\pm$,
      $ K^\pm \etapr$, and
      $ \omega \pi^\pm$.  
      With the full
      CLEO II and CLEO II.V datasets 
      statistical precision on \acp\ is in the range of $\pm 0.12$ to
      $\pm 0.25$ depending on the mode. 
      All quoted results are preliminary.
\end{abstract}
\newpage
\begin{center}
T.~E.~Coan,$^{1}$ V.~Fadeyev,$^{1}$ I.~Korolkov,$^{1}$
Y.~Maravin,$^{1}$ I.~Narsky,$^{1}$ R.~Stroynowski,$^{1}$
J.~Ye,$^{1}$ T.~Wlodek,$^{1}$
M.~Artuso,$^{2}$ R.~Ayad,$^{2}$
% ...
\end{center}
 
\small
\begin{center}
$^{1}${Southern Methodist University, Dallas, Texas 75275}\\
$^{2}${Syracuse University, Syracuse, New York 13244}\\
% ...
\end{center}

\newpage

\section{Introduction}
CP violating phenomena arise in the Standard Model because of the
single complex parameter in the quark mixing matrix.\cite{buras} Such
phenomena are expected to occur widely in $B$ meson decays and are the
incentive for most of the current $B$-physics initiatives in the
world. As of yet there is little direct experimental
evidence.  CDF's recent determination\cite{cdf} of $\sin2\beta$ at the $2\sigma$
level, which followed upon earlier less sensitive searches by both
CDF\cite{CDFearly} and OPAL\cite{opal}, is consistent with expectations for mixing-induced
Standard Model CP violation and to date the only evidence of CP
effects in $B$ mesons.  

Direct CP violation however is also
anticipated to play a prominent role in the CP phenomena of
$B$ decay.  To date the only published search for direct CP
violation in $B$ decay is the recent CLEO limit\cite{bsgacp} on \acp\ in
$b\to s\gamma$.  CP asymmetries may show up in any 
mode where there are two or more participating
diagrams which differ in weak and strong phases. $B\to K\pi$ modes,
for instance, involve $b\to u$ tree diagrams carrying the weak phase
${\rm Arg}(V^*_{ub}V_{us})\equiv \gamma$ and $b\to s$ penguin diagrams
carrying the weak phase ${\rm Arg}(V^*_{tb}V_{ts}) = \pi$.  Though the
branching ratios are small, such cases are experimentally
straightforward to search for.  Rate differences between $B\to K^+\pi^-$
and $\bar B\to K^-\pi^+$ decays would be unambiguous signals of 
direct CP violation
if seen.

We report here five searches for direct CP violation in charmless
hadronic $B$ decay modes, based on the full CLEO II and CLEO II.V
datasets which together comprise 9.66 million $B\bar B$ events.
The modes searched for are the three $K\pi$ modes,
$ K^\pm \pi^\mp$,
$ K^\pm \pi^0$,
$ \KS \pi^\pm$,
the mode $ K^\pm \etapr$, and the vector-pseudoscalar mode
$ \omega \pi^\pm$. In all but the first case the flavor of the parent $b$ or
$\bar b$ quark is tagged simply by the sign of the 
high momentum charged hadron; for 
$K^\pm\pi^\mp$ one must further identify $K$ and $\pi$. 

In what follows we will refer when needed to the generic final state
from $b$ and $\bar b$ as $f$ and $\bar f$ respectively. The
corresponding event yields of signal and background
we will label as $\sig$, $\sigbar$, \bkg, and \bkgbar.  
We define the sign of \acp\
with the following convention:
\begin{equation}
\acp \equiv 
\frac{\branch(b\to f)-\branch(\bbar\to\fbar)}
{\branch(\bbar\to\fbar)+\branch( b\to f)}
~=~
\frac{\sig - \sigbar}{\sig + \sigbar}
\label{eqn:acpdef}
\end{equation}

The statistical precision one can achieve in a measurement of \acp\ depends 
on the signal yield $S\equiv\sig+\sigbar$, the 
CP-symmetric background $B=\bkg + \bkgbar$, any correlation
$\chi$ that might exist between the measurements of $f$ and \fbar,
and of course on \acp\ itself:
\begin{equation}
\sigma^2_{\acp} =
\frac{1-\acp^2}{S}
\left(
1 +
\frac{B}{S}
\left(\frac{1+\acp^2}{1-\acp^2}\right)
-2\chi\sqrt{
\frac{1-\acp^2}{4} + 
\frac{B^2}{S^2} + 
\frac{B}{S}}
\right)
\label{eqn:acpstats}
\end{equation}
In most cases there is no chance of confusing $f$ and \fbar\ so
$\chi=0$. However for $f=K^-\pi^+$ and $\fbar = K^+\pi^-$, a small
degree of crossover is possible due to imperfect particle identification;
we find $\chi = -0.11$ for this case. For $\chi\sim 0$, $B/S\sim 1$,
and $\acp\sim 0$
one has an easy rule of thumb, $\sigma\approx\sqrt{2/S}$.  For $S\sim 100$
this means one expects statistical precision in the 
neighborhood of $\pm 0.15$.  As will be discussed later, systematic
errors are small, and consequently the precision in \acp\ measurements can
be expected to be dominated by statistical errors for a long time
to come.  As can be seen in Eq. \ref{eqn:acpstats} the statistical
error is in turn dominated by the leading $1/\sqrt{S}$ coefficient
which reminds us that the only path to better \acp\ measurements will be
more data. Improvements to analysis technique that reduce $B/S$ or $\chi$
have less impact.

\section{Theoretical Expectations}
The existence of a CP violating rate asymmetry depends on having
both two different CP nonconserving weak phases and 
two different CP conserving strong phases.  The former may arise
from either the Standard Model CKM matrix or from new physics, while
the latter may arise from the absorptive part of a penguin diagram
or from final state interaction effects.  The difficulty of calculating
strong interaction phases, particularly when long distance non-perturbative
effects are involved, largely precludes reliable predictions.  
Under well-defined model assumptions, however, numerical estimates 
may be made and the dependence on both model parameters and CKM
parameters can be probed. A recent and comprehensive review of CP
asymmetries under the assumption of generalized factorization has
been published by Ali {\it et al} \cite{ali}. We quote in Table
\ref{tbl:ali} their predictions for the modes examined in this paper.

\begin{table}[ht]
\begin{center}
\begin{tabular}{lc}
\hline
Mode & \acp\  \cr
\hline
$B\to K^\pm\pi^\mp$ & $ 0.037 \to 0.106 $ \cr
$B\to K^\pm\pi^0$   & $ 0.026 \to 0.092 $ \cr
$B\to K^0\pi^\pm$   & $ 0.015$            \cr
$B\to K^\pm\etapr$  & $ 0.020 \to 0.061 $ \cr
$B\to \omega\pi^\pm$& $ -0.120 \to 0.024$ \cr
\hline
\end{tabular}
\caption{CP asymmetry predictions. 
The range of \acp\ values
reflects a range of model parameters.  
CKM parameters $\rho$ and
$\eta$ are set to $\rho=0.12$, $\eta=0.34$. 
Taken from Ali $et~al$, Ref. 6, with signs changed to
match the convention of Eq. 1.}
\label{tbl:ali}
\end{center}
\end{table} 

If final state interactions are not neglected, however, strong phases
as large as $90^\circ$ are not ruled out and $|\acp|$ could reach
0.44 in favorable cases.\cite{someone} In this
nonperturbative regime numerical predictivity is limited, but a
variety of relationships among asymmetries or $f,\fbar$ rate
differences can be found in the literature.\cite{allthose}

\section{Data Set, Detector, Event Selection}
The data set used in this analysis was collected with the CLEO II and
CLEO II.V detectors at the Cornell Electron Storage Ring (CESR).  It
consists of $9.1~{\rm fb}^{-1}$ taken at the $\Upsilon$(4S)
(on-resonance) and $4.5~{\rm fb}^{-1}$ taken below $B\bar{B}$
threshold.  The below-threshold sample is used for continuum
background studies.  The on-resonance sample contains 9.66~million
$B\bar{B}$ pairs.  This is a factor 2.9 increase in the number of
$B\bar{B}$ pairs over the published measurements of the modes 
considered here \cite{kpiPRL},\cite{etaprPRL},\cite{omegaPRL}. In
addition, the CLEO II.V data set, which has significantly improved
particle identification and momentum resolution as compared with CLEO
II, now dominates the data set.

 CLEO II and CLEO II.V are general purpose solenoidal magnet
detectors, described in detail elsewhere~\cite{detector}.  In CLEO II,
the momenta of charged particles are measured in a tracking system
consisting of a 6-layer straw tube chamber, a 10-layer precision drift
chamber, and a 51-layer main drift chamber, all operating inside a 1.5
T superconducting solenoid.  The main drift chamber also provides a
measurement of the specific ionization loss, $dE/dx$, used for
particle identification.  For CLEO II.V the 6-layer straw tube chamber
was replaced by a 3-layer, double-sided silicon vertex detector, and
the gas in the main drift chamber was changed from an argon-ethane to
a helium-propane mixture.  Photons are detected using 7800-crystal
CsI(Tl) electromagnetic calorimeter. Muons are identified using
proportional counters placed at various depths in the steel return
yoke of the magnet.

Charged tracks are required to pass track quality cuts based on the
average hit residual and the impact parameters in both the $r-\phi$
and $r-z$ planes.  Candidate \KS\ are selected from pairs of tracks
forming well-measured displaced vertices.  Furthermore, we require the
\KS\ momentum vector to point back to the beam spot and the
$\pi^+\pi^-$ invariant mass to be within $10$~MeV, two standard
deviations ($\sigma$), of the \KS\ mass.  Isolated showers with
energies greater than $40$~MeV in the central region of the CsI
calorimeter and greater than $50$~MeV elsewhere, are defined to be
photons.  Pairs of photons with an invariant mass within 
2.5$\sigma$ of the nominal $\pi^0$ ($\eta$) mass are kinematically fitted 
with the mass constrained to the nominal $\pi^0$ ($\eta$) mass.  To reduce
combinatoric backgrounds we require the lateral shapes of the showers
to be consistent with those from photons.  To suppress further low
energy showers from charged particle interactions in the calorimeter
we apply a shower-energy-dependent isolation cut.

Charged particles are identified as kaons or pions using $dE/dx$.
Electrons are rejected based on $dE/dx$ and the ratio of the track
momentum to the associated shower energy in the CsI calorimeter.  We
reject muons by requiring that the tracks do not penetrate the steel
absorber to a depth greater than seven nuclear interaction lengths.
We have studied the $dE/dx$\ separation between kaons and pions for
momenta $p \sim 2.6$~GeV$/c$\ in data using 
$D^0\rightarrow K^- \pi^+(\pi^0)$ decays; we find a separation of $(1.7\pm
0.1)~\sigma$ for CLEO II and $(2.0\pm 0.1)~\sigma$ for CLEO II.V.

Resonances are reconstructed through the decay channels: 
$\eta'\to\eta\pi^+\pi^-$ with $\eta\to\gamma\gamma$; $\eta'\to\rho\gamma$
with $\rho\to\pi^+\pi^-$; and $\omega\to\pi^+\pi^-\pi^0$.

\section{Analysis}
The \acp\ analyses presented are intimately related to the
corresponding branching ratio determinations presented in separate
contributions to this conference.\cite{br} We summarize here the
main points of the analysis.

We select hadronic events and impose efficient quality cuts
on tracks, photons, $\pi^0$ candidates, and \KS\ candidates.
We calculate a beam-constrained $B$ mass $M = \sqrt{E_{\rm b}^2 -
p_B^2}$, where $p_B$ is the $B$ candidate momentum and $E_{\rm b}$ is
the beam energy.  The resolution in $M$\ ranges from 2.5 to 3.0~${\rm
MeV}/{\it c}^2$, where the larger resolution corresponds to the
$B^\pm\to h^\pm\pi^0$ decay.  We define $\Delta E = E_1 + E_2 - E_{\rm b}$,
where $E_1$ and $E_2$ are the energies of the daughters of the $B$
meson candidate.  The resolution on $\Delta E$ is mode-dependent.  For
final states without photons the $\Delta E$ resolution for CLEO
II.V(II) is $20(26)$~MeV.  Most other modes are only slightly worse
but  for the $B^\pm\to h^\pm \pi^0$ analysis, the $\Delta E$ resolution 
is worse by about a factor of two and
becomes asymmetric because of energy loss out of the back of the CsI
crystals.  The energy constraint also helps to distinguish between
modes of the same topology.  For example, $\Delta E$ for $B
\rightarrow K^+ \pi^-$, calculated assuming $B \rightarrow
\pi^+\pi^-$, has a distribution that is centered at $-42$~MeV, giving
a separation of $2.1(1.6)\sigma$ between $B \rightarrow K^+ \pi^-$ and
$B \rightarrow \pi^+\pi^-$ for CLEO II.V(II).  We accept events with
$M$\ within $5.2-5.3$~$\rm {GeV/c^2}$\ and $|\Delta E|<200$~MeV.  The
$\Delta E$ requirement is loosened to 300 MeV for the $B^\pm\to h^\pm
\pi^0$ analysis.  This
fiducial region includes the signal region, and a sideband for
background determination.  Similar regions are included around each of 
the resonance masses ($\eta'$, $\eta$, and $\omega$) in the likelihood
fit.  For the $\eta'\to\rho\gamma$ case, the $\rho$ mass is not included
in the fit; we require $0.5\gev < m_{\pi\pi} < 0.9\gev$.

We have studied backgrounds from $b\to c$\ decays and other $b\to u$\
and $b\to s$\ decays and find that all are negligible for the analyses
presented here. The main background arises from $e^+e^-\to q\bar q$\
(where $q=u,d,s,c$).  Such events typically exhibit a two-jet
structure and can produce high momentum back-to-back tracks in the
fiducial region.  To reduce contamination from these events, we
calculate the angle $\theta_{sph}$ between the sphericity axis\cite{shape}
of the candidate tracks and showers and the sphericity axis of the
rest of the event. The distribution of $\cossph$\ is strongly
peaked at $\pm 1$ for $q\bar q$\ events and is nearly flat for $B\bar
B$\ events. We require $|\cossph|<0.8$\ which eliminates $83\%$\
of the background.  For \etapr\ and $\omega$ modes the cut is made at 0.9.

Additional discrimination between signal and $q\bar q$\ background is
provided by a Fisher discriminant technique as described in detail in
Ref.~\cite{bigrare}.  The Fisher discriminant is a linear combination
${\cal F}\equiv \sum_{i=1}^{N}\alpha_i y_i$\ where the coefficients
$\alpha_i$ are chosen to maximize the separation between the signal
and background Monte-Carlo samples.  The 11 inputs, $y_i$, are
$|\cos\theta_{cand}|$ (the cosine of the angle between the candidate
sphericity axis and beam axis), the ratio of Fox-Wolfram moments
$H_2/H_0$~\cite{fox}, and nine variables that measure the scalar sum
of the momenta of tracks and showers from the rest of the event in
nine angular bins, each of $10^\circ$, centered about the candidate's
sphericity axis.  For the \etapr\ and $\omega$ modes, $|\cos\theta_B|$
(the angle between the $B$ meson momentum and beam axis), is used
instead of $H_2/H_0$.

Using a detailed GEANT-based Monte-Carlo
simulation~\cite{geant} we determine overall detection efficiencies
of $15-46\%$, as listed in
Table~\ref{tbl:inputs}. Efficiencies contain secondary branching fractions for
$K^0\to \KS\to \pi^+\pi^-$\ and $\pi^0\to \gamma\gamma$ 
as well as $\etapr$ and $\omega$ decay modes where
applicable.  We estimate a systematic error on the efficiency using
independent data samples.

In Table \ref{tbl:inputs} we summarize, for each mode, cuts, efficiencies, and
the total number of events which pass the cuts and
enter the likelihood fit described in the next paragraph.

\begin{table}[ht]
\begin{center}
\begin{tabular}{lcccccc}
\hline
              & $K^\pm \pi^0$  & $K^0\pi^\pm$   & \op\     & \multicolumn{2}{c}{\ek\ } & \kp\  \cr
              &         &         &          & \epp\    & \rg\    &         \cr
\hline
\mb\          & 5.2-5.3 & 5.2-5.3 & 5.2-5.3  & 5.2-5.3  & 5.2 5.3 & 5.2-5.3 \cr
$|\de|$       & $<0.3$  & $<0.2$  & $<0.2$   & $<0.2$   & $<0.2$  & $<0.2$  \cr
$|\cossph|$   & $<0.8$  & $<0.8$  & $<0.9$   & $<0.9$   & $<0.9$  & $<0.8$  \cr
Efficiency    & $0.40$  & $0.15$  & $0.26$   & $0.27$   & $0.29$  & $0.46$  \cr
Events in Fit &  6991   & 1558     & 21337   &  395     & 10284   & 5407    \cr 
\hline
\end{tabular}
\caption{Summary of cuts and number of events used in each mode. }
\label{tbl:inputs}
\end{center}
\end{table}

To extract signal and background yields we perform unbinned
maximum-likelihood (ML) fits using $\Delta E$, $M$, ${\cal F}$,
$|\cos\theta_B|$ (if not used in ${\cal F}$),
and $dE/dx$ (where applicable), daughter resonance mass (where
applicable), and helicity angle in the daughter decay (where
applicable).  
The free parameters to be fitted are the 
asymmetry ($(f-\fbar)/(f+\fbar)$) and the sum ($f+\fbar$)
in both signal and background. 
In most cases there is more than one possible signal component
and its corresponding background component,
as for instance we fit simultaneously for \kpz\ and \ppz\ to
ensure proper handling of the $K\pi$ identification information. 
The probability distribution functions (\pdf s)
describing the distribution of events in each variable are
parametrized by simple forms (gaussians, polynomials, etc.) whose
parameter values are determined in separate studies.  For signal \pdf\
shapes the parameter determinations are made by fitting signal Monte
Carlo events.  Backgrounds in these analyses are dominated by
continuum \eeqq\ events, and we determine parameters of the background
\pdf s by fitting data taken below the $\Upsilon(4S)$ resonance or
data taken on resonance but lying in the sidebands of the signal
region.  The uncertainties associated with such fits are
used later to assess the final systematic error.

%-------------------------------------------------------------------------
%
\section{Results}
%
%-------------------------------------------------------------------------

%\subsection{Mode by mode analysis}

\begin{figure}[htbp]
\begin{center}
\epsfig{file=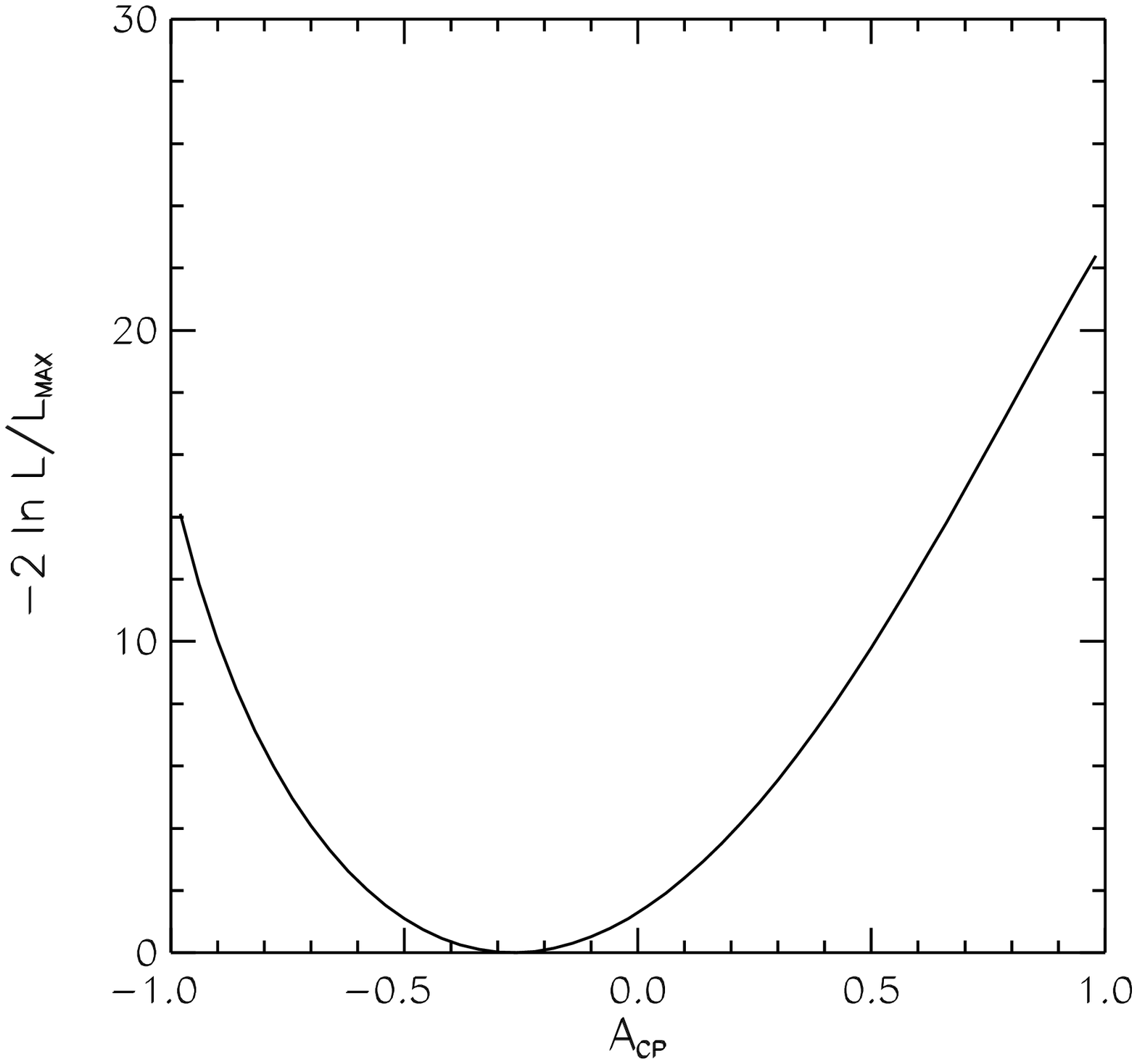,width=3.5in}
\caption{$B\to K^\pm\pi^0$ mode: $-2\ln(\like/\like_{max})$ as a function of \acp.}
\label{fig:kpz_asymmetry}
\end{center}
\end{figure}

%-------------------------------------------------------------------------
%
\paragraph{($K\pi^0$)}
%
%-------------------------------------------------------------------------
In the mode $B^\pm\to K^\pm\pi^0$ we find a total of $\aerr{45.6}{11.2}{10.2}$
events with an asymmetry of $\acp(K\pi^0)=-0.27\pm 0.23$. This corresponds
to 
$28.9\pm 7.5$ $K^+\pi^0$ and
$16.8\pm 7.5$ $K^-\pi^0$ events.
(Here and elsewhere we will quote the yields \sig\ and \sigbar\
for the convenience of the reader. These are 
derivative quantities as the fit directly extracts \acp\ and
$\sig +\sigbar$.)
We note that $\pi\pi^0$ which
is analyzed simultaneously but does not have a sufficiently
significant yield to measure a branching fraction shows an
asymmetry of $\acp(\pi\pi^0) = -0.03 \pm 0.40$. 
Fig. \ref{fig:kpz_asymmetry}  shows the likelihood function
dependence on $\acp(K\pi^0)$. 
As a cross check we measure the asymmetry of the background events,
finding $0.023\pm 0.026$ for $K\pi^0$ background and 
$0.000\pm 0.017$ for $\pi\pi^0$ background. These values are consistent
with the expected null result for continuum background.

%-------------------------------------------------------------------------
%
\paragraph{($K_s\pi$)}
%
%-------------------------------------------------------------------------
In the mode $B^\pm\to K_s\pi^\pm$ we find a total of $\aerr{25.2}{6.4}{5.6}$
events with an asymmetry of $\acp(K_s\pi)=0.17\pm 0.24$. This corresponds
to 
$10.2\pm 4.0$ $K_s\pi^+$ and
$14.5\pm 4.4$ $K_s\pi^-$ events.
The background events show an asymmetry of $-0.02\pm 0.04$, consistent with
zero as expected.

\begin{figure}[htbp]
\begin{center}
\epsfig{file=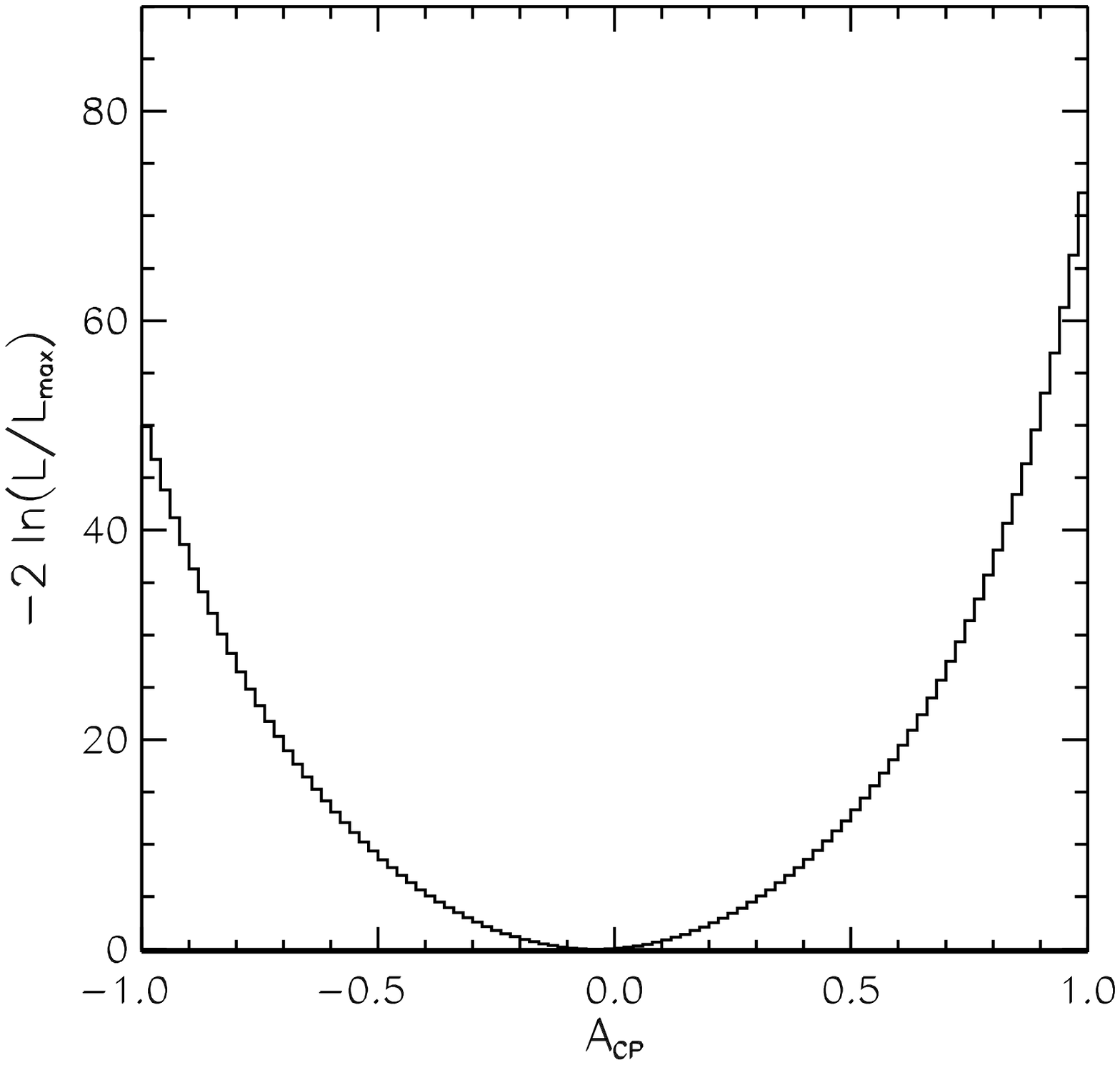,width=3.5in,clip=,
                             bburx=525, bbury=485,
        bbllx=20, bblly=15}
\caption{$K^\pm\pi^\mp$ mode: $-2\ln(\like/\like_{max})$ plotted versus \acp.}
\label{fig:kpi_asymmetry}
\end{center}
\end{figure}

%-------------------------------------------------------------------------
%
\paragraph{($K\pi$)}
%
%-------------------------------------------------------------------------
In the mode $B\to K^\pm\pi^\mp$ we find a total of $\aerr{80.2}{11.8}{11.0}$
events with an asymmetry of $\acp(K\pi)=-0.04\pm 0.16$. This corresponds
to 
$\aerr{41.6}{8.9}{8.0}$ $K^+\pi^-$ and
$\aerr{38.6}{9.0}{8.1}$ $K^-\pi^+$ events.
The dependence of the likelihood function on \acp\ is shown
in  Fig. \ref{fig:kpi_asymmetry}.
The background events show an asymmetry of $-0.02\pm 0.04$, consistent
with zero as expected.

\begin{figure}[htbp]
\begin{center}
\epsfig{file=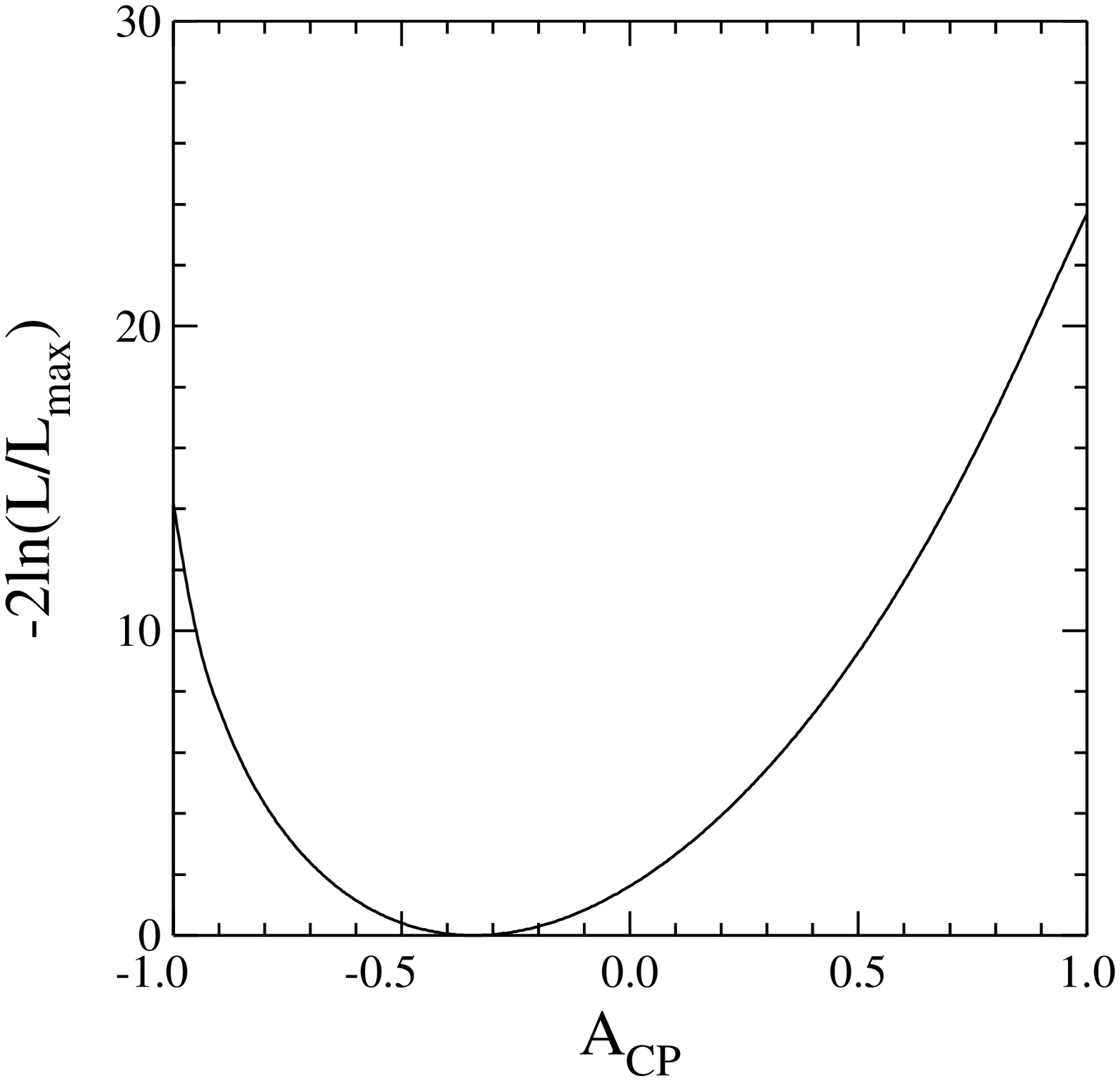,width=3.5in,clip=,
                             bburx=525, bbury=515,
        bbllx=2, bblly=15}
\caption{$B^\pm\to\op$ mode: $-2\ln(\like/\like_{max})$ as a function of \acp.}
\label{fig:opi_asymmetry}
\end{center}
\end{figure}

%-------------------------------------------------------------------------
%
\paragraph{($\omega\pi$)}
%
%-------------------------------------------------------------------------
In the mode $B^\pm\to\op$ we find a total of $\aerr{28.5}{8.2}{7.3}$ events
with $\acp(\omega\pi) = -0.34 ^{+0.24} _{-0.26}$. 
This corresponds to 
$\aerr{19.1}{6.8}{5.9}$ $\omega\pi^+$ events and
$\aerr{9.4}{4.9}{4.0}$ $\omega\pi^-$ events.
Figure~\ref{fig:opi_asymmetry}
shows the likelihood as a function of \acp.
The fit also allows for a possible charge asymmetry
in the continuum background, and finds values consistent with zero:
$-0.013\pm 0.015$ for $\omega K^\pm$ background, and 
$-0.001\pm 0.010$ for $\omega \pi^\pm$ background. 

\begin{figure}[htbp]
\begin{center}
\epsfig{file=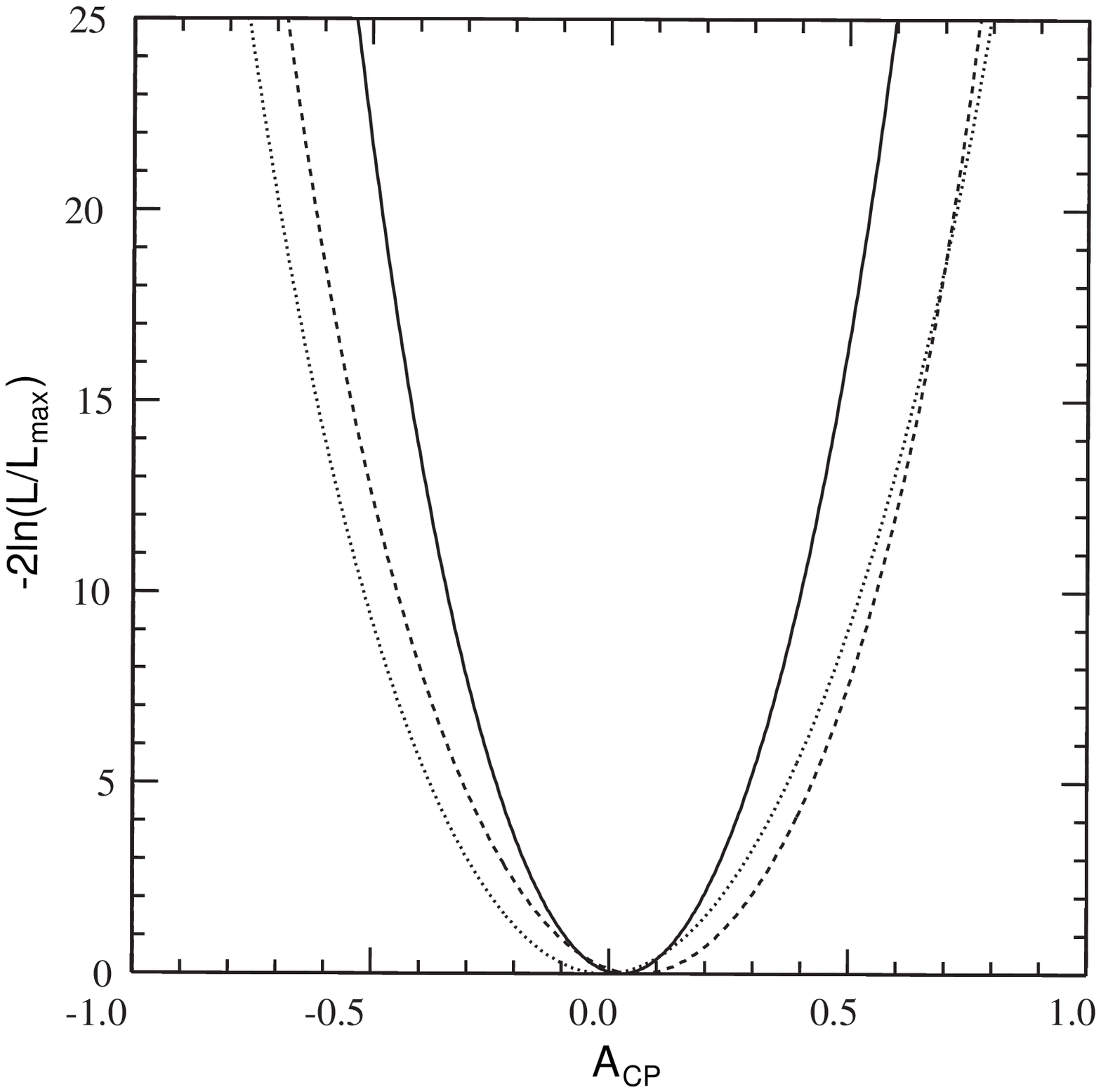,width=3.5in}
\caption{$\etapr K^\pm$ mode: $-2\ln(\like/\like_{max})$ plotted versus \acp. The
dashed curve shows the fit result for the $\etapr\to\eta\pi\pi$ decay,
the dotted curve for $\etapr\to\rho\gamma$ and the solid curve is the
combined result.}
\label{fig:eklike}
\end{center}
\end{figure}

%-------------------------------------------------------------------------
%
\paragraph{($\etapr K$)}
%
%-------------------------------------------------------------------------
We find $\acp(\etapr K) = 0.03\pm 0.12$.  Separating the $\etapr K$ signal
sample by submodes $\etapr \to \epp$ and $\etapr\to\rg$ we find $\acp = 0.06\pm 0.17$ and
$\acp = -0.01\pm 0.17$, respectively. Fig. \ref{fig:eklike} shows the
dependence of the fitted likelihood function on \acp\ for the
separate and combined submodes.
Background $\etapr K$ events are found to have an asymmetry of
$-0.01\pm 0.07$ in the \epp\ mode, and
$-0.009\pm 0.015$ in the \rg\ mode, both consistent with zero as expected.

%-------------------------------------------------------------------------
%
\section{Systematic Errors}
%
%-------------------------------------------------------------------------

The charge asymmetries measured in this analysis hinge primarily on
the properties of high momentum tracks.  The charged meson that tags
the parent $b/\bar b$ flavor has momentum in all cases between 2.3 and
2.8 GeV/c.  In independent studies using very large samples of high
momentum tracks we have searched for and set stringent limits on the
extent of possible charge-correlated bias in the CLEO detector and
analysis chain for tracks in the $2-3$GeV range.  Based on a sample of
8 million tracks, we find $\acp$ bias introduced by differences in
reconstruction efficiencies for positive and negative high momentum
tracks passing the same track quality requirements as are imposed in
this analysis is less than $\pm 0.002$.  For $K^\pm\pi^\mp$
combinations where differential charge-correlated efficiencies must
also be considered in correlation with $K/\pi$ flavor, we use 37,000
$D^0\to K\pi(\pi^0)$ decays and set a corresponding limit on \acp\
bias at $\pm 0.005$. These $D^0$ decays, together with an additional
24,000 $D^\pm_{(s)}$ decays, are also used to set a tight upper limit
of 0.4 MeV/c on any charge-correlated or charge-strangeness-correlated
bias in momentum measurement. The resulting limit on \acp\ bias from
this source is $\pm 0.002$.  We conclude that there is no significant
\acp\ bias introduced by track reconstruction or selection.  We note
that for each mode we crosscheck the asymmetry of the background
events (normally a fairly large sample) and find results consistent
with zero as anticipated.

\begin{figure}[htbp]
\begin{center}
\epsfig{file=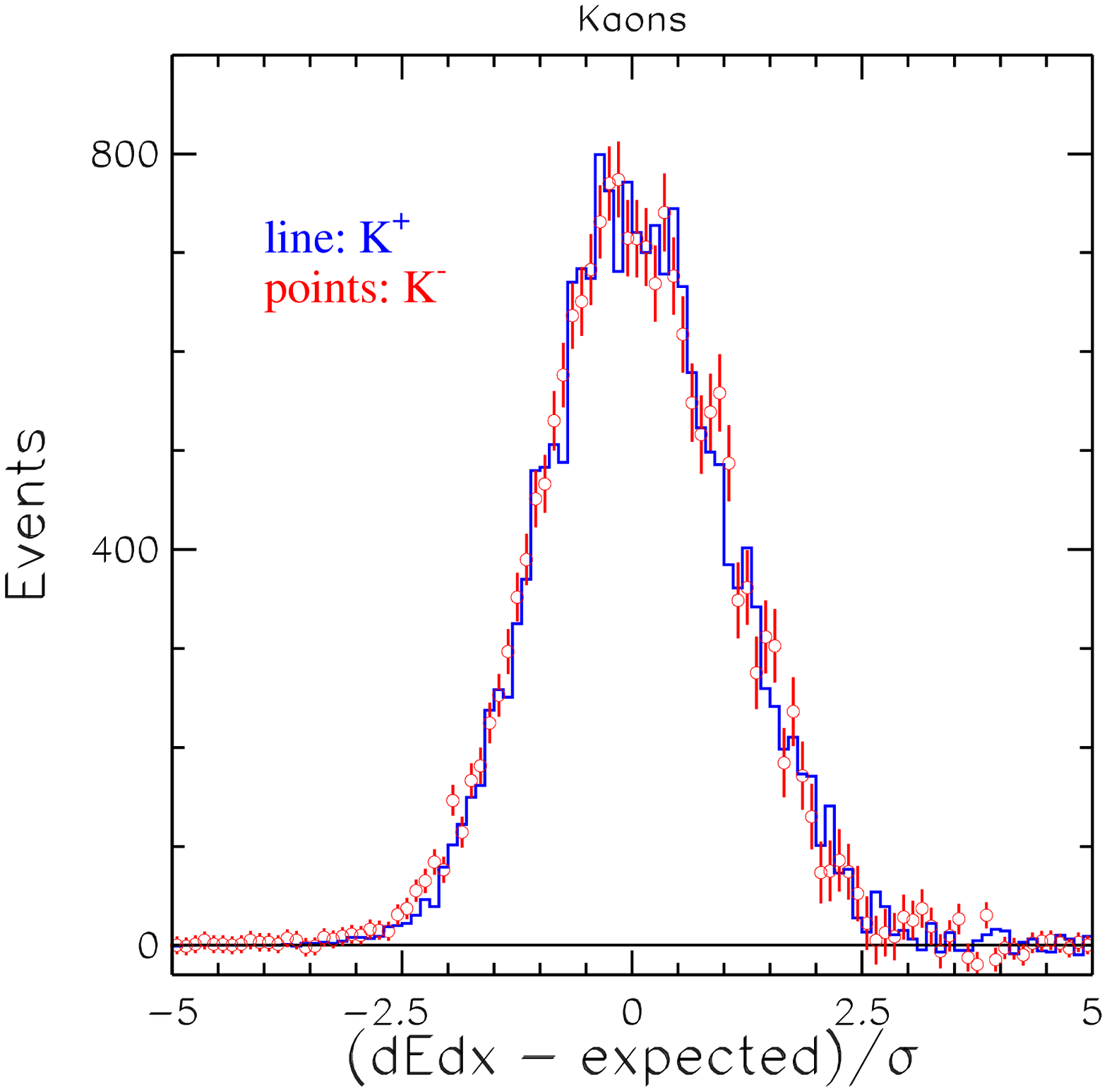,width=2.7in}
\epsfig{file=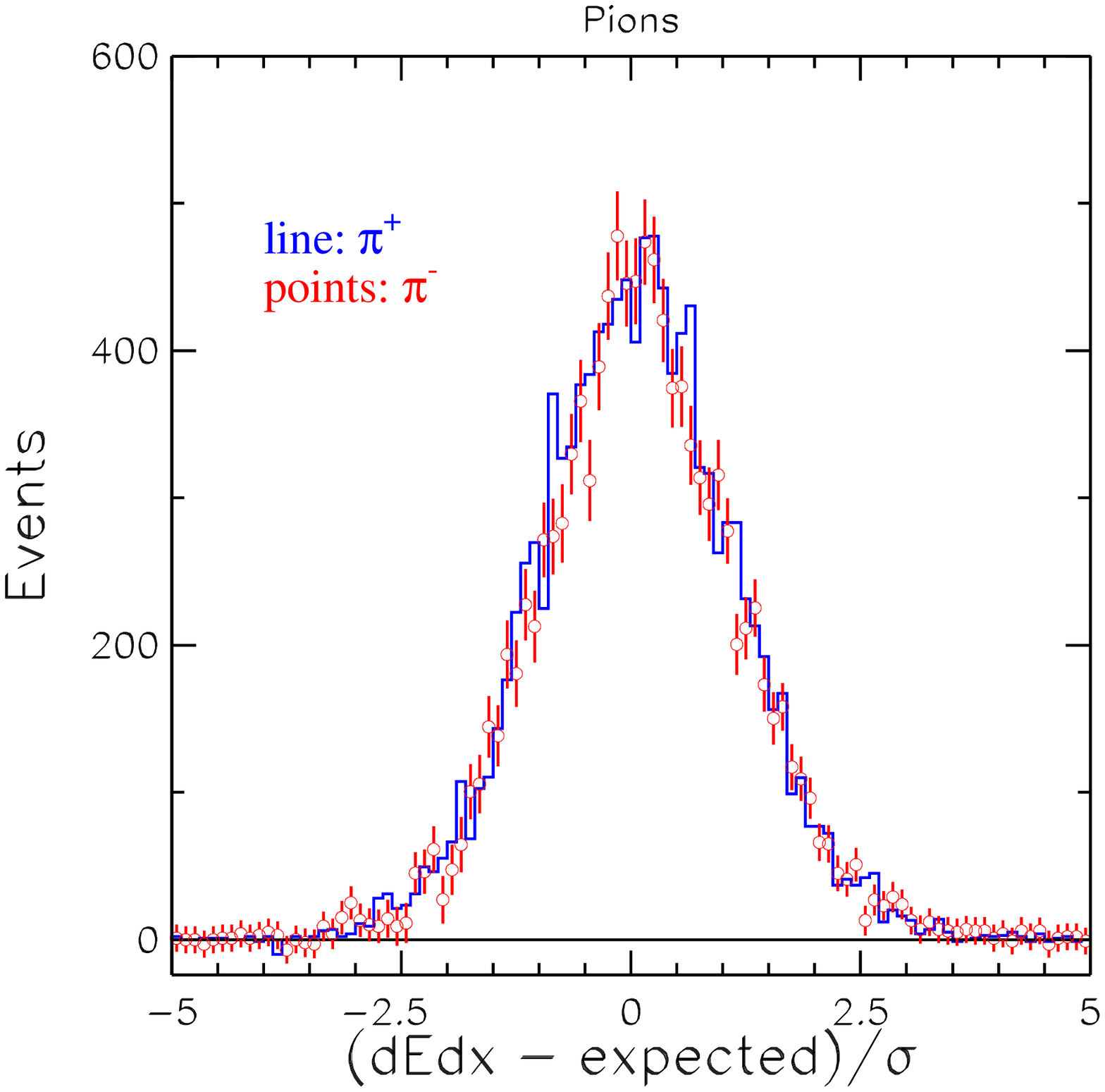,width=2.7in}
\caption{Normal central \dedx\ distributions for $K^\pm$ and $\pi^\pm$.}
\label{fig:dedx}
\end{center}
\end{figure}

Particle identification information for $K^\pm$ and $\pi^\pm$ is
shown in Fig. \ref{fig:dedx}. No significant differences are seen
between different charge species. Quantification of the effect
on \acp\ is covered by \pdf\ variation studies discussed
immediately below.

All \pdf\ shapes which are used in the maximum likelihood fits are
varied within limits prescribed by the fits which determine the shape
parameters to assess the systematic error associated with
uncertainty in the parameters.  The resulting changes in \acp\ are
summed in quadrature to estimate the systematic error due to possible
misparametrization of the \pdf\ shapes. This contribution
to the systematic error ranges from $\pm 0.02$ to $\pm 0.04$ 
depending on mode.

We choose to assign a conservative systematic error $\pm 0.05$ for all
modes.

\vskip 0.5cm
\begin{table}[ht]
\begin{center}
\begin{tabular}{|lcccc|}
\noalign{\hrule\vskip 0.15truecm}
Mode &  $\sig$              & $\sigbar$            & $\acp$                  & 90\% CL          \cr
     &                      &                      &                         & interval         \cr
\noalign{\vskip 0.15truecm\hrule\vskip 0.15truecm}
$K\pi^0$         &  $16.8\pm 7.5$      & $28.9\pm 7.5$        & $-0.27\pm 0.23\pm 0.05$          & $[-0.70, 0.16]$   \cr
$K^0_S\pi$       & $14.5 \pm 4.4$      & $10.2 \pm 4.0$       & $0.17\pm 0.24\pm 0.05$           & $[-0.27, 0.61]$   \cr
$K\pi$           &$38.6_{-8.1}^{+9.0}$ &$41.6_{-8.0}^{+8.9}$  & $-0.04\pm 0.16\pm 0.05$          & $[-0.35, 0.27]$   \cr
$\etapr K$       & $51.7\pm 9.2$       & $48.7\pm 8.9$        & $0.03\pm 0.12\pm 0.05$           & $[-0.22, 0.28]$  \cr
$\omega\pi$     &$\aerr{9.4}{4.9}{4.0}$&$\aerr{19.1}{6.8}{5.9}$& $-0.34\pm 0.25\pm 0.05$         & $[-0.80, 0.12]$  \cr
\hline
\end{tabular}
\caption{\acp\ measurements in five charmless $B$ decay modes. CLEO 1999 Preliminary.}
\label{tbl:sum}
\end{center}
\end{table}

\begin{figure}[htbp]
\begin{center}
\epsfig{file=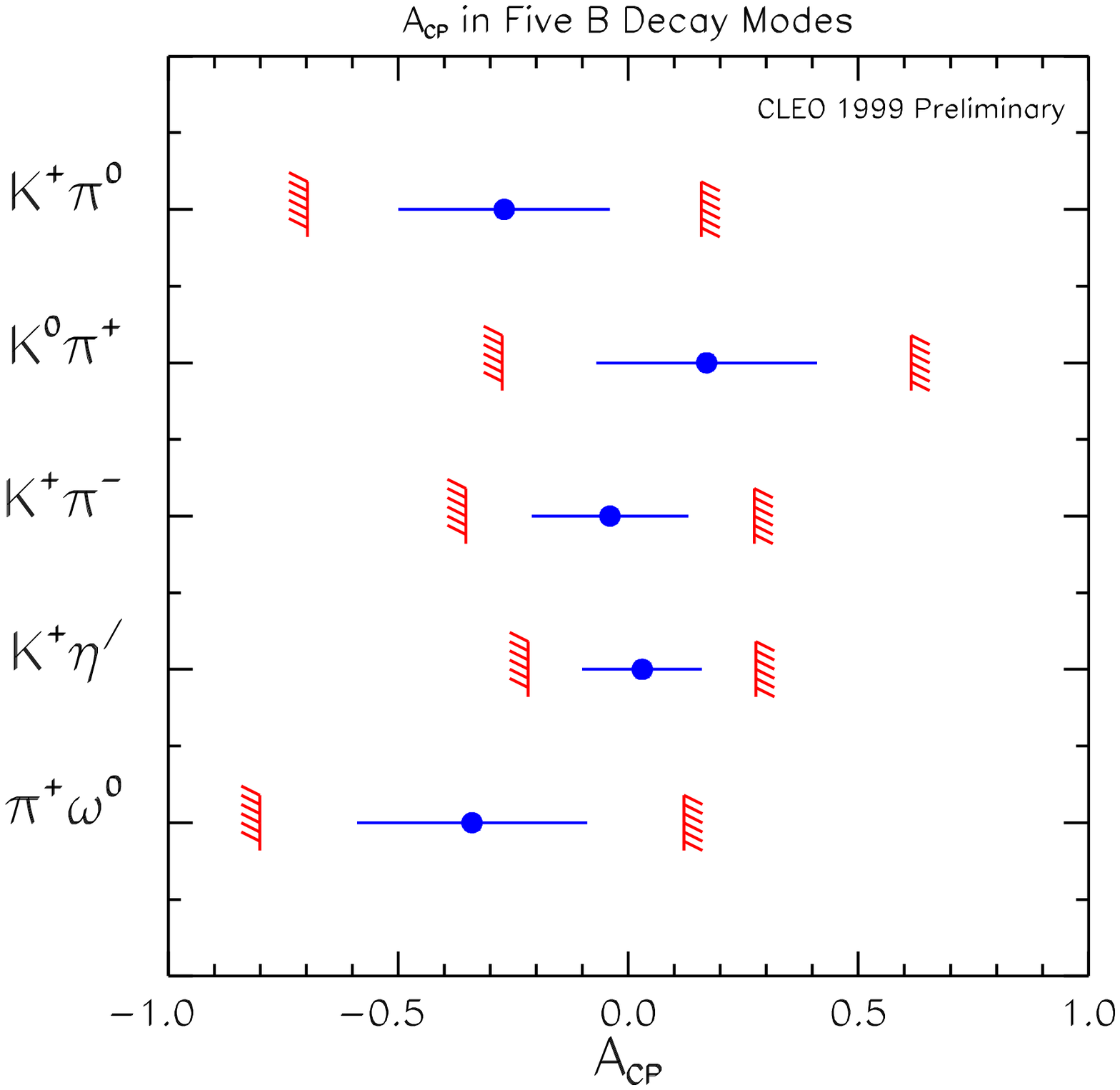,width=5.0in}
\caption{Results for \acp\ in five
charmless $B$ decay modes.
The data points show the central value with $1\sigma$ error bars; hatching
indicates the range of the 90\% CL interval.}
\label{fig:acp_intervals}
\end{center}
\end{figure}

\section{Summary}
Table \ref{tbl:sum} and Fig. \ref{fig:acp_intervals} 
summarize all results. 

We have presented in this paper the first measurements of charge
asymmetries in hadronic $B$ decay.  We see no evidence for CP
violation in the five modes analyzed here and set 90\% CL intervals
(systematics included) that reduce the possible range of \acp\ by as
much as a factor of four.  While the sensitivity is not yet sufficient
to address the rather small \acp\ values predicted by factorization
models, extremely large \acp\ values which could arise if large phases
were available from final state interactions are firmly ruled out. For
the case of $K\pi$ and $\etapr K$ we can exclude at 90\% confidence
values of $|\acp|$ greater than 0.35 and 0.28 respectively.

The search for CP violating asymmetries will intensify with the new
datasets expected in the coming years. The precision of such searches
will be statistics limited and should improve with integrated
luminosity as $1/\sqrt{\lum}$. If modes with large asymmetries 
and reasonable branching ratios exist
they could be found within a few years.

\end{document}